\documentclass{osa-article}
\journal{osac}
\articletype{Research Article}

\usepackage{booktabs}
\usepackage{footnote}
\makesavenoteenv{tabular}
\usepackage{rotating}

\usepackage{siunitx}
\usepackage{amsmath}

\begin{document}

\title{Spectral noise in quantum frequency down-conversion from the visible to the telecommunication C-band}

\author{Peter C.~Strassmann, Anthony Martin, Nicolas Gisin, Mikael Afzelius\authormark{*}}

\address{Group of Applied Physics, University of Geneva, CH-1211 Geneva 4, Switzerland}

\email{\authormark{*}mikael.afzelius@unige.ch}

\begin{abstract*}
We report a detailed study of the noise properties of a visible-to-telecom photon frequency converter based on difference frequency generation (DFG).
The device converts \SI{580}{\nano\meter} photons to \SI{1541}{\nano\meter} using a strong pump laser at \SI{930}{\nano\meter}, in a periodically poled lithium niobate ridge waveguide.
The converter reaches a maximum device efficiency of \SI{46}{\percent} (internal efficiency of \SI{67}{\percent}) at a pump power of \SI{250}{\milli\watt}.
The noise produced by the pump laser is investigated in detail by recording the noise spectra both in the telecom and visible regimes, and measuring the power dependence of the noise rates.
The noise spectrum in the telecom is very broadband, as expected from previous work on similar DFG converters.
However, we also observe several narrow dips in the telecom spectrum, with corresponding peaks appearing in the \SI{580}{\nano\meter} noise spectrum.
These features are explained by sum frequency generation of the telecom noise at wavelengths given by the phase matching condition of different spatial modes in the waveguide.
The proposed noise model is in good agreement with all the measured data, including the power-dependence of the noise rates, both in the visible and telecom regime.
These results are applicable to the class of DFG converters where the pump laser wavelength is in between the input and target wavelength.
\end{abstract*}

\section{Introduction}
\label{sec:introduction}

Quantum frequency conversion (QFC) from the visible domain to the telecommunication (telecom) bands plays an important role in the development of fiber-based quantum networks.
This is because several matter systems that are currently under development as quantum nodes emit photons in the visible domain, while optical fibers have minimum losses in the telecom bands.
These matter systems include nitrogen-vacancy (NV) centers in diamond (\SI{637}{\nano\meter}) \cite{Hensen2015,Humphreys2018}, europium- (\SI{580}{\nano\meter}) \cite{Laplane2017} and praseodymium-doped (\SI{606}{\nano\meter}) \cite{Ferguson2016,Maring_2017a} rare-earth (RE) crystals, which emit in the yellow-red wavelength range.
Other examples include trapped Yb$^{+}$ (\SI{370}{\nano\meter}) \cite{Moehring2007,Vittorini2014}, Ba$^{+}$ (\SI{493}{\nano\meter}) \cite{Crocker2018,Siverns2019} and Sr$^{+}$ (\SI{422}{\nano\meter}) \cite{Wright2018} single ions, which emit in the near-UV and blue regions.
Here we are specifically interested in the conversion of photons emitted by europium-doped crystals at \SI{580}{\nano\meter}.

A convenient technique for achieving QFC of these wavelengths into the telecom bands is to use a single-stage difference frequency generation (DFG) process in a non-linear $\chi^{(2)}$ medium.
This requires a strong pump laser at $1/\lambda_{\mathrm{pump}} = 1/\lambda_{\mathrm{vis}} - 1/\lambda_{\mathrm{tele}}$ where $\lambda_{\mathrm{vis}}$ is the wavelength of the visible photon to be converted and $\lambda_{\mathrm{tele}}$ is the target wavelength in a telecom band.
If we target the telecom C-band around \SI{1550}{\nano\meter}, the pump laser will be in the $\lambda_{\mathrm{pump}} = \SIrange[range-phrase=-]{900}{1100}{\nano\meter}$ region for the NV and RE systems \cite{Dreau_2018,Maring_2018a}, and in the range of $\lambda_{\mathrm{pump}} = \SIrange[range-phrase=-]{480}{720}{\nano\meter}$ for Yb$^{+}$, Ba$^{+}$ and Sr$^{+}$ ions \cite{Ruetz_2017,Wright2018}, if we assume the operating wavelengths given above.

DFG in quasi-phase-matched non-linear crystal waveguides can reach high conversion efficiencies \cite{Pelc2010,Zaske2012,Ikuta_2014a,Maring_2018a,Dreau_2018,Ruetz_2017}, an important feature of practical QFC.
But an intense pump laser with a wavelength in between the input and target wavelengths generates noise at the target wavelength.
This is due to non-phase-matched broadband spontaneous parametric down conversion (SPDC) of the pump laser \cite{Pelc2010}.
This noise contribution scales linearly with the pump power, as observed and verified in numerous experiments \cite{Pelc2010,Tamura_2018a,Maring_2018a,Dreau_2018}.
Yet, recent experiments have shown that the noise can be suppressed to acceptable levels by strong filtering around the target wavelength \cite{Maring_2017a,Maring_2018a,Dreau_2018,Ruetz_2017,Wright2018}.
Indeed, the noise is very broadband (typically \SI{>100}{\nano\meter}), such that the amount of photon noise per spectral/temporal mode is much less than $1$.
For some of the systems cited above this might require filtering down to the system bandwidth of \SIrange{1}{10}{\mega\hertz}.

The telecom SPDC noise can also be converted back to the input wavelength by phase-matched sum frequency generation (SFG) where $1/\lambda_{\mathrm{vis}} = 1/\lambda_{\mathrm{pump}} + 1/\lambda_{\mathrm{tele}}$.
This cascaded process results in a noise rate, at $\lambda_{\mathrm{vis}}$, that scales quadratically with the pump power, as also observed in numerous experiments \cite{Thew2006,Ma2012,Maring2014,Ruetz_2017}.
As the SFG is phase-matched, only telecom noise photons within the SFG bandwidth should be efficiently converted to the visible and one would expect to observe a narrowband dip in the broadband noise at $\lambda_{\mathrm{tele}}$.
Simultaneously a narrowband source of noise should appear at $\lambda_{\mathrm{vis}}$, as illustrated in Fig.~\ref{fig:setup}a.
Such a narrowband noise peak was observed by Rutz \textit{et al.}~\cite{Ruetz_2017}, with the approximate spectral width of the SFG.
One would also expect a sub-linear power dependence of the noise in the telecom regime, due to the SFG conversion of noise photons, provided that the telecom noise is filtered with a bandwidth smaller than the SFG bandwidth.
This was recently observed by Maring \textit{et al.}~\cite{Maring_2018a}, which is an indirect confirmation that there is a dip in the broadband telecom noise, although the dip was not confirmed by a direct measurement of the noise spectrum.
One of the goals of this article is to study the telecom and visible noise spectra in detail, and correlate these to the power dependence of the noise in these parts of the spectra.

Other QFC experiments based on DFG have been performed of input photons in the deep red or near-infrared range \cite{Zaske2011,Pelc2012,Zaske2012,Fernandez-Gonzalvo_2013,Kuo2018,Weber2018}, with the target in a telecom band, which are closely related to the experiments discussed here.
However, in terms of the noise mechanism these experiments are different.
In some cases the pump was spectrally close to the target wavelength \cite{Zaske2011,Fernandez-Gonzalvo_2013}, such that Raman noise induced by the pump was the dominant source of noise.
In other experiments the pump had a wavelength well above the target wavelength \cite{Pelc2012,Zaske2012,Kuo2018,Weber2018}, which strongly suppresses the SPDC noise.
Similarly Esfandyarpour \textit{et al.}~demonstrated a cascaded, two-stage DFG conversion of \SI{650}{\nano\meter} photons to the telecom C-band using a pump at \SI{2.2}{\micro\meter} \cite{Esfandyarpour2018}.
The noise properties presented here apply specifically to the QFC of visible photons where the pump is spectrally well separated from the target wavelength with $\lambda_{\mathrm{vis}} < \lambda_{\mathrm{pump}} < \lambda_{\mathrm{tele}}$.

In this article we describe a single-stage DFG device for achieving QFC of photons emitted by europium-doped quantum nodes at $\lambda_{\mathrm{vis}} = \SI{580}{\nano\meter}$, into the telecom C-band at $\lambda_{\mathrm{vis}} = \SI{1541}{\nano\meter}$ using a pump laser at $\lambda_{\mathrm{pump}} = 930 \mathrm{\ nm}$.
The device is based on a ridge waveguide on a periodically poled lithium niobate (PPLN) crystal.
We present a detailed characterization of the noise spectrum in both the telecom band and the visible region around \SI{580}{\nano\meter}.
We observe dips in the otherwise broadband telecom noise, each associated to the SFG phase matching wavelength of different spatial modes in the waveguide.
In the visible noise spectrum, peaks are observed at the corresponding wavelengths given by energy conservation.
The SPDC and SFG processes generate the telecom and visible noise rates deviating from linear and quadratic scaling, respectively, at high pump powers.
We reach a maximum external device efficiency of \SI{46}{\percent} (including coupling through the waveguide), with an internal waveguide efficiency of \SI{67}{\percent}.
The spectral noise rate at the \SI{1541}{\nano\meter} target wavelength is \SI{5}{\hertz\per\watt\per\centi\meter} in a \SI{1}{\mega\hertz} bandwidth, in the linear regime at low pump powers, which is on par with the lowest noise rate reached in similar DFG experiments \cite{Dreau_2018}.

\section{Experimental set-up}
\label{sec:experiment}

\begin{figure}[!t]
\centering
\includegraphics[width=12.5cm]{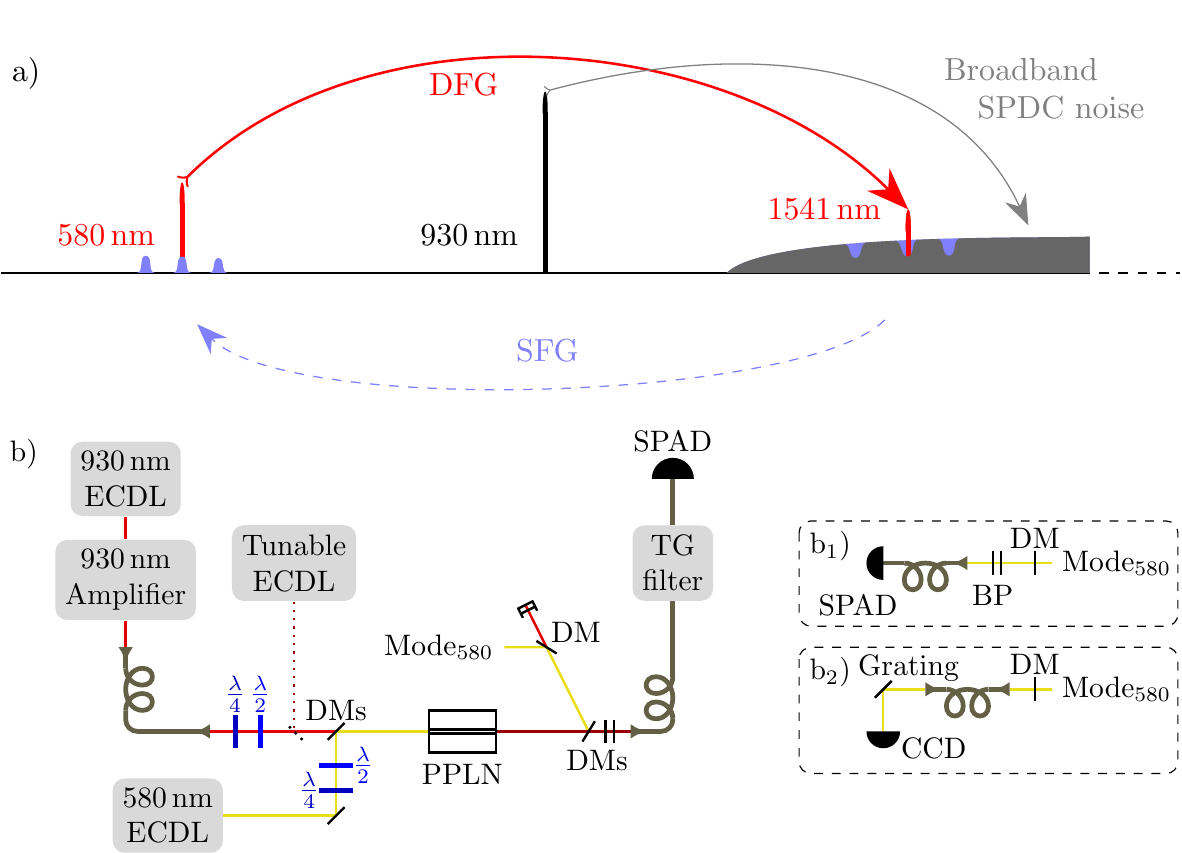}
\caption{Implementation of the frequency conversion.
(a) Overview of the DFG conversion process from \SI{580}{\nano\meter} to \SI{1541}{\nano\meter}, together with the different pump-induced noise processes.
The noise processes are the non-phase-matched SPDC by the strong pump beam at \SI{930}{\nano\meter} and the phase-matched SFG of part of the SPDC noise into the visible domain around \SI{580}{\nano\meter}.
The SFG process occurs for the fundamental spatial mode at a wavelength of \SI{1541}{\nano\meter}, but also for higher-order spatial modes at other wavelengths.
(b) Schematic of the experimental setup.
The pump light at \SI{930}{\nano\meter} is amplified and its spatial mode is cleaned by a single-mode fiber.
The pump beam is overlapped with the laser beams at \SI{580}{\nano\meter} and \SI{1541}{\nano\meter} using dichroic mirrors (DMs).
Quarter ($\lambda/4$) and half ($\lambda/2$) waveplates align the polarizations to the vertical axis of the PPLN waveguide.
All laser beams are coupled into the PPLN waveguide.
At the output they are again separated by DMs and directed to different setups for the noise and efficiency measurements, as explained in detail in Section \ref{sec:experiment}.
ECDL = external-cavity diode laser, SPAD = single-photon avalanche diode, TG = tunable grating filter from JDS Uniphase (TB9226), BP = band-pass filter.
}
\label{fig:setup}
\end{figure}

The experimental set-up is shown in Fig.~\ref{fig:setup}b.
The core of the experiment is a ridge waveguide on a PPLN crystal manufactured by NTT.
The PPLN waveguide has a length of \SI{40}{\milli\meter} and a cross-section of \SI{12.7}{\micro\meter} by \SI{10.7}{\micro\meter}, with a poling period of \SI{9.19}{\micro\meter}.
A Peltier element stabilizes the waveguide at a temperature of \SI{53}{\celsius}, as required to reach the quasi-phase-matching for a type 0 non-linear optical process with the involved wavelengths.
Both waveguide facets are anti-reflection coated at the three wavelengths involved in the DFG, in order to avoid any etalon effects and to maximize the transmission.

The pump laser is an external-cavity diode laser (ECDL) at \SI{930}{\nano\meter}, which is amplified by a tapered amplifier diode to about \SI{1.3}{\watt}.
The spatial mode was not in a Gaussian TEM$_{00}$ mode, hence a single mode \SI{930}{\nano\meter} fiber was used as a spatial mode cleaner.
This allowed maximizing the power in the fundamental mode of the waveguide.
It also prevented intensity hotspots on the input facet of the waveguide, thereby reducing the risk of damaging its surface.
The mode cleaner reduced the maximum pump power before the waveguide guide to at most \SI{510}{\milli\watt}, of which about \SI{80}{\percent} was coupled into the waveguide.
The coupled pump power varied slightly between experiments (\SI{\pm5}{\percent}) and it was calibrated for each measurement.

To characterize the DFG conversion process, in particular its efficiency, we used another ECDL laser at \SI{580}{\nano\meter}.
Also, for the spectral measurements of the corresponding SFG process, a tunable telecom ECDL laser was employed.
All three input beams were overlapped using dichroic mirrors (DMs), and their linear polarization were aligned to the vertical axis of the waveguide.

At the output all three wavelengths were spectrally separated using DMs.
The telecom mode was coupled into a single mode fiber (\SI{75}{\percent} coupling efficiency) and passed through a tunable grating (TG) filter with a bandwidth of \SI{200}{\pico\meter} (\SI{25}{\giga\hertz}) and a transmission of about \SI{40}{\percent}.
For the noise measurements the telecom photons were detected by a free-running InGaAs single-photon avalanche diode (SPAD)  (efficiency \SI{10}{\percent} and dark count rate \SI{340}{\hertz}), while for the DFG conversion efficiency measurements the light was detected by a linear photodiode.
The tunability of the TG filter allowed measuring the noise rate over a large spectral range (\SIrange{1520}{1575}{\nano\meter}).

The \SI{580}{\nano\meter} output mode was analyzed using different set-ups.
We measured the noise spectrum using a home-made spectrometer based on a grating and a CCD camera, see inset b$_2$ in Fig.~\ref{fig:setup}b.
It has a measured instrumental resolution of (FWHM) \SI{130(30)}{\pico\meter} (or \SI{116(27)}{\giga\hertz}).
As will be discussed in Section \ref{sec:results}, the noise spectrum consists of discrete peaks where the peak around the input wavelength of \SI{580}{\nano\meter} is of special interest.
This noise contribution was measured as a function of pump power by filtering the mode with a \SI{580}{\nano\meter} band-pass (BP) filter and detecting the photons with a free-running silicon SPAD (efficiency \SI{56}{\percent} at \SI{580}{\nano\meter} and dark count rate \SI{70}{\hertz}), as shown in inset b$_1$ in Fig.~\ref{fig:setup}b.
Finally, we also measured the SFG spectrum in order to identify higher order spatial modes in the waveguide, by tuning the telecom laser and recording the SFG signal at \SI{580}{\nano\meter} with a linear photo diode (not shown in Fig.~\ref{fig:setup}b).

\section{Experimental results}
\label{sec:results}

As discussed in Section \ref{sec:introduction}, the SFG plays an important role for the noise in QFC experiments based on the DFG process, see Fig.~\ref{fig:setup}a.
We therefore start by presenting and discussing the SFG characterization in Section \ref{sec:results:SFG_mode_analysis}, which also allowed us to identify higher order modes and their phase matching wavelengths.
In Section \ref{sec:results:spec_analysis} we present the noise spectrum measurements in the spectral regions centered at \SI{580}{\nano\meter} and \SI{1541}{\nano\meter}.
In Section \ref{sec:results:power_dep} we present the measurements of the power dependence of the DFG conversion efficiency and of the noise rate, which are correlated to measured noise spectra.

\subsection{Sum Frequency Generation and spatial mode characterization}
\label{sec:results:SFG_mode_analysis}

\begin{figure}[!b]
\centering
\includegraphics{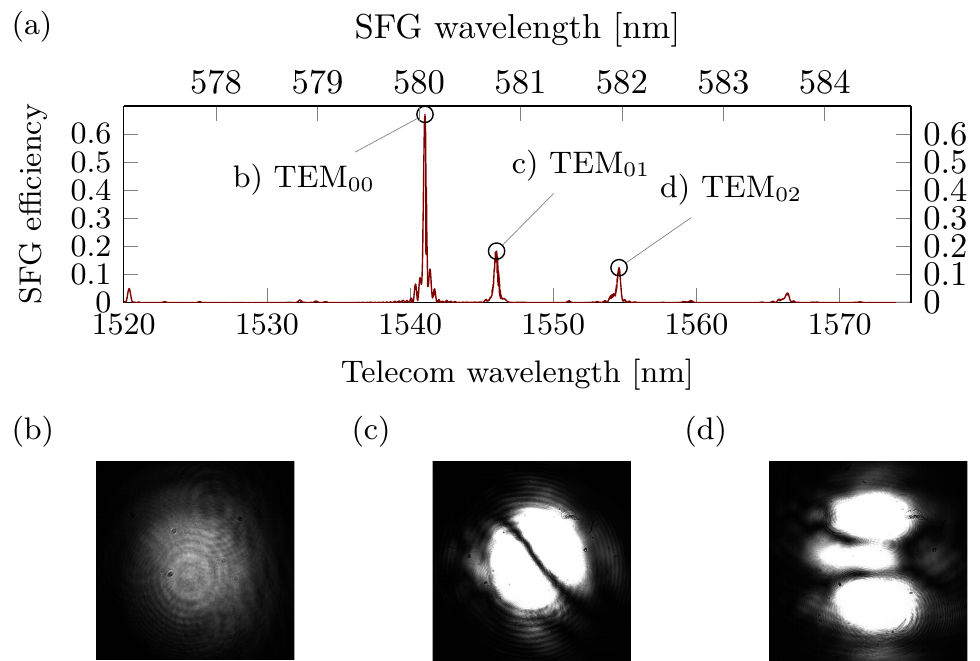}
\caption{Sum frequency generation (SFG) measurement.
(a) SFG signal efficiency as a function of the telecom laser wavelength for a coupled pump power of \SI{440}{\milli\watt}.
The vertical scale shows the external SFG conversion efficiency, defined in the same way as the DFG conversion efficiency discussed in Section \ref{sec:results:power_dep}.
The horizontal top scale shows the corresponding SFG wavelengths for a pump wavelength of \SI{930}{\nano\meter}.
The three main SFG modes were imaged onto a CCD camera, from which the following spatial modes could be identified:
(b) TEM$_{00}$ at $\SI{1541.0}{\nano\meter} \widehat{=} \SI{580.0}{\nano\meter}$,
(c) TEM$_{01}$ at $\SI{1546.0}{\nano\meter} \widehat{=} \SI{580.7}{\nano\meter}$, and
(d) TEM$_{02}$ at $\SI{1554.6}{\nano\meter} \widehat{=} \SI{581.9}{\nano\meter}$.
}
\label{fig:result_SFG}
\end{figure}

The SFG spectrum was recorded by scanning the telecom laser from \SIrange{1520}{1575}{\nano\meter} with the \SI{930}{\nano\meter} pump laser at its full power, while the \SI{580}{\nano\meter} laser was blocked.
The SFG signal was detected with a free-space linear photodiode placed after the DMs that spectrally seperated the beams after the waveguide, see Fig.~\ref{fig:setup}b.

The experimental spectrum is shown in Fig.~\ref{fig:result_SFG}a, where one observes a strong SFG signal with the telecom laser at \SI{1541.0}{\nano\meter}, \SI{1546.0}{\nano\meter} and \SI{1554.6}{\nano\meter}, corresponding to the SFG wavelengths of \SI{580.0}{\nano\meter}, \SI{580.7}{\nano\meter} and \SI{581.9}{\nano\meter}, respectively.
The spatial modes of the SFG signal at these wavelengths were imaged onto a CCD camera, which allowed us to identify them as being the TEM$_{00}$, TEM$_{01}$ and TEM$_{02}$ Hermite-Gaussian modes, as shown in Fig.~\ref{fig:setup}b-d.

\subsection{Spectral noise measurements}
\label{sec:results:spec_analysis}

\begin{figure}[!b]
\centering
\includegraphics{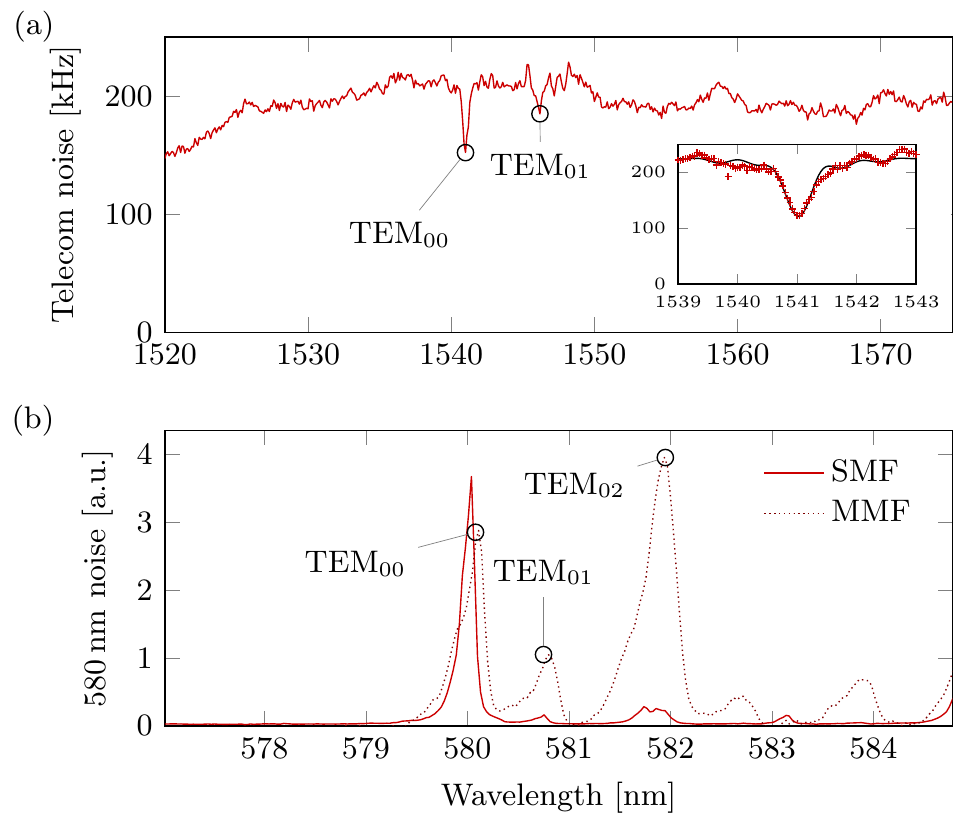}
\caption{
Spectrally resolved noise measurements at a coupled pump power of \SI{440}{\milli\watt}.
(a) The broadband noise spectrum at the telecom C-band shows dips due to the SFG conversion of noise photons in the TEM$_{00}$ and TEM$_{01}$ spatial modes identified in Fig.~\ref{fig:result_SFG}.
The TG filter scan had a step size of \SI{.1}{\nano\meter}.
Each data point represents the average count rate integrated over \SI{10}{\second}.
The slight drop in count rate below \SI{1525}{\nano\meter} is due to the transmission profile of one of the DMs.
Inset: Higher resolution scan of the dip at \SI{1541}{\nano\meter} using a TG step size of \SI{.04}{\nano\meter}.
(b) The noise spectrum recorded in the \SI{580}{\nano\meter} range recorded with the spectrometer using a CCD integration time of \SI{300}{\second}.
The \SI{580}{\nano\meter} output from the waveguide was coupled into either a single- (solid line) or multi-mode (dashed line) fiber, which explains the difference in sensitivity to higher order spatial modes.
The spectral resolution is limited by the resolution of the home-made spectrometer.
}
\label{fig:result_spectrum}
\end{figure}

The noise at the single photon level was recorded by injecting the pump laser into the waveguide, while blocking the \SI{580}{\nano\meter} and telecom laser, and measuring the photon rate at the output of the waveguide.
In the telecom region the noise spectrum was recorded by moving the TG filter stepwise and measuring the photon rate at each step using the InGaAs SPAD.
The visible noise spectrum around \SI{580}{\nano\meter} was recorded with the home-made spectrometer coupled to the CCD camera.

The telecom noise spectrum recorded between \SIrange[range-phrase={ and }]{1520}{1575}{\nano\meter} is shown in Fig.~\ref{fig:result_spectrum}a.
The observed noise spectrum is essentially flat over the entire bandwidth, as expected from a non-phase-matched SPDC process \cite{Pelc2010}.
But one can clearly observe two dips in the spectrum, at the wavelengths of \SI{1541}{\nano\meter} and \SI{1546}{\nano\meter}.
The \SI{1541}{\nano\meter}-dip is due to the SFG in the fundamental TEM$_{00}$ spatial mode of the waveguide, while the \SI{1546}{\nano\meter}-dip is due to the TEM$_{01}$ spatial mode as identified in Fig.~\ref{fig:result_SFG}.
A higher resolution noise spectrum around the \SI{1541}{\nano\meter}-dip is shown in the inset of Fig.~\ref{fig:result_spectrum}a.

The associated noise spectrum around \SI{580}{\nano\meter} is shown in Fig.~\ref{fig:result_spectrum}b.
The output mode at \SI{580}{\nano\meter} was either coupled into a single- (SMF) or multi-mode (MMF) fiber before entering the spectrometer, see Fig.~\ref{fig:setup}b$_2$.
The MMF fiber accepts higher order spatial modes at the output of the waveguide, resulting in several strong SFG noise peaks.
Based on the SFG spectrum in Fig.~\ref{fig:result_SFG}, these can be easily identified as resulting from the SFG conversion of the telecom noise in the spatial SFG modes TEM$_{00}$, TEM$_{01}$ and TEM$_{02}$.
The SMF fiber, on the other hand, strongly suppresses the higher order modes and the noise spectrum is dominated by the SFG in the fundamental TEM$_{00}$ mode.

The noise and SFG spectra presented in Fig.~\ref{fig:result_SFG} and Fig.~\ref{fig:result_spectrum} show excellent overall agreement.
The data clearly supports a noise model dominated by pump-induced broadband SPDC in the telecom range, but where SFG conversion of the noise into the visible range strongly reduces the noise at wavelengths corresponding to different spatial modes supported by the waveguide.
In the fundamental TEM$_{00}$ SFG mode at \SI{1541}{\nano\meter} we observe a significant noise reduction of about \SI{40}{\percent} at the highest pump power.
The \SI{1541}{\nano\meter}-dip shown in the inset of Fig.~\ref{fig:result_spectrum} has a Gaussian linewidth of \SI{540(40)}{\pico\meter}, which corresponds to a linewidth of \SI{500(40)}{\pico\meter} after deconvolution with the TG filter linewidth of \SI{200(20)}{\pico\meter}.
This is twice as wide as the \SI{230(10)}{\pico\meter} linewidth of SFG peak of the TEM$_{00}$ mode measured in Fig.~\ref{fig:result_SFG}a.
We believe this is due to the shorter effective SFG interaction length experienced by the telecom noise photons, which are created throughout the entire waveguide length.
This will also reduce the effectiveness of the SFG.
In the next section the power dependence of the noise will be studied and modeled in more detail.

\subsection{Noise rate as a function of pump power}
\label{sec:results:power_dep}

\begin{figure}[!t]
\centering
\includegraphics{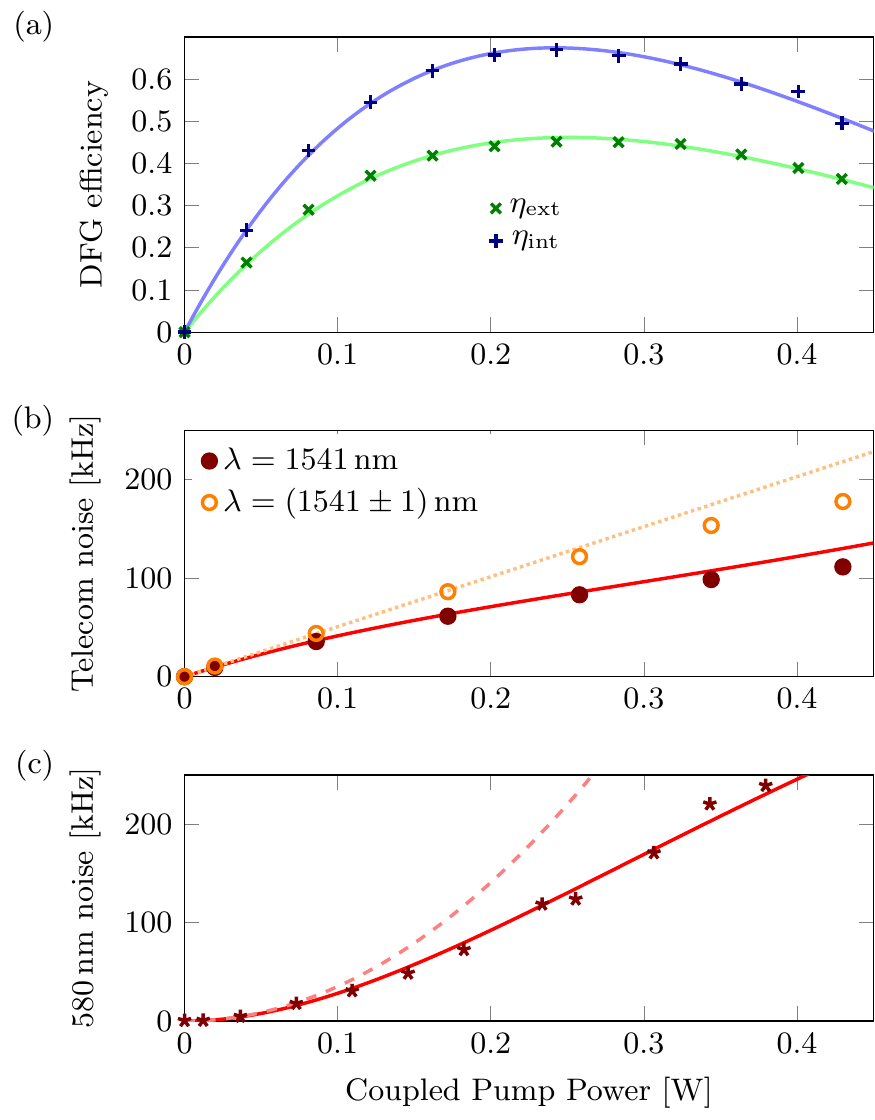}
\caption{Power dependence of the DFG efficiency and noise rates normalized to the output of the waveguide.
(a) External, $\eta_\text{ext}$, and internal, $\eta_\text{int}$, DFG conversion efficiencies.
The solid lines represents fits to Eq.~\ref{eq:eta_power}.
(b) Telecom noise rate measured through the \SI{200}{\pico\meter} TG filter tuned to the phase-matching wavelength of \SI{1541}{\nano\meter} (solid circles), centered on the dip shown in the inset of Fig.~\ref{fig:result_spectrum}a, or detuned by \SI{1}{\nano\meter} from this wavelength (open circles).
The detuned data points represent the average noise rate measured at \SI{\pm1}{\nano\meter} from the \SI{1541}{\nano\meter} dip.
The dashed line is a linear fit of the first four points for the detuned case.
The solid line represent a model taking into account the SFG noise reduction, as detailed in Section \ref{sec:results:power_dep}.
(c) Noise rate around \SI{580}{\nano\meter} measured with the waveguide output coupled to a SMF fiber (cf.~Fig.~\ref{fig:result_spectrum}b).
The lines represent the models detailed in Section \ref{sec:results:power_dep}.%
}
\label{fig:result_power}
\end{figure}

The external DFG conversion efficiency, $\eta_\text{ext}$, can be obtained by measuring the $\lambda_\text{in}=\SI{580}{\nano\meter}$ laser power before the waveguide ($P^{\lambda_\text{in}}_\text{in}$), and the converted $\lambda_\text{out}=\SI{1541}{\nano\meter}$ telecom power after the waveguide ($P^{\lambda_\text{out}}_\text{out}$), as a function of the injected pump power.
We further express the conversion efficiency in terms of photon rates, in which case the external efficiency is calculated using $\eta_\text{ext}=(P^{\lambda_\text{out}}_\text{out}/P^{\lambda_\text{in}}_\text{in})(\lambda_\text{out}/\lambda_\text{in})$.
The internal DFG conversion efficiency, $\eta_\text{int}$, can be measured by comparing the \SI{580}{\nano\meter} laser output power without ($P^{\lambda_\text{in}}_\text{outref}$) and with ($P^{\lambda_\text{in}}_\text{out}$) the pump laser, thus we measure the relative depletion of the \SI{580}{\nano\meter} light due to the DFG conversion.
The internal conversion efficiency can then be expressed as $\eta_\text{int}=1-P^{\lambda_\text{in}}_\text{out}/P^{\lambda_\text{in}}_\text{outref}$.

In Fig.~\ref{fig:result_power}a, the internal and external DFG conversion efficiencies are plotted as a function of the coupled \SI{930}{\nano\meter} pump laser power, up to the full power of \SI{440}{\milli\watt}.
The maximum efficiencies $\eta_\text{max}$ of \SI{67}{\percent} (internal) and \SI{46}{\percent} (external) are reached for a pump power of \SI{250}{\milli\watt}.
The external efficiency is lower due to in-out coupling losses and waveguide propagation losses, while the internal conversion efficiency we believe is limited by the spatial mode matching of the three highly non-degenerate wavelengths involved in the DFG (\SIlist{580;930;1541}{\nano\meter}).

According to theory the efficiency of the DFG should be described by the formula \cite{Langrock2005},
\begin{equation}
\eta(P_p)=\eta_\text{max}\sin^2\left(L\sqrt{\eta_\text{n} P_p}\right),
\label{eq:eta_power}
\end{equation}
where $L$ is the length of the non-linear medium, $P_p$ is the injected pump power, and $\eta_\text{n}$ is a conversion parameter characteristic to the specific device.
Both the internal and external conversion efficiencies can be fitted using a common $\eta_\text{n}$ parameter, see the solid lines in Fig.~\ref{fig:result_power}a, yiedling $\eta_\text{n} = \SI{63(2)}{\percent\per\watt\per\centi\meter^2}$.

The noise rates in the telecom region were measured through the TG filter at the single photon level using the InGaAs SPAD, cf.~Fig.~\ref{fig:setup}b.
All the rates given here are normalized with the known transmission losses, such that they represent the photon rate at the output of the waveguide.
The TG filter was either centered on the DFG/SFG phase matching wavelength of \SI{1541}{\nano\meter}, or detuned by \SI{\pm1}{nm} from this wavelength.
At \SI{1541}{\nano\meter} the noise rate increases with the pump power, as shown in Fig.~\ref{fig:result_power}b, but the dependence is less strong than the linear dependence (see dashed line in Fig.~\ref{fig:result_power}b) that would be expected from a purely SPDC dominated noise process.
This sub-linear dependence is due to the noise reduction caused by the SFG process, as also observed by Maring \textit{et al.}~\cite{Maring_2018a}.
The effect of the SFG process is more clearly seen by comparing with the TG filter detuned by \SI{\pm1}{nm} from the DFG/SFG phase matching wavelength, where the dependence becomes almost linear as shown in Fig.~\ref{fig:result_power}b.
We note that the amount of noise reduction seen at full power is consistent with the spectral noise dip observed in Fig.~\ref{fig:result_spectrum}a.

In Ref.~\cite{Maring_2018a} Maring \textit{et al.}~proposed a model to explain the power dependence of the noise rate,
\begin{align}
R_\text{tele}(P_p)
&=\alpha_N P_p \int_0^L \left(1-\eta_\text{max}\sin^2\left(x\sqrt{\eta_\text{n} P_p}\right)\right) \, \mathop{}\!\mathrm{d}x\\
&=\alpha_N P_p L \left(1-\frac{\eta_\text{max}}{2}\left(1-\frac{\sin\left(2L\sqrt{\eta_\text{n} P_p}\right)}{2L\sqrt{\eta_\text{n} P_p}}\right)\right),
\label{eq:Maring}
\end{align}
where the last explicit expression is given in the present work.
Here $\alpha_N$ is the device specific noise parameter and without the SFG noise reduction the noise rate would ideally scale linearly as $\alpha_N P_p L$.
The SFG causes a sub-linear dependence due to the second term in the parenthesis, which only depends on parameters already determined from the DFG efficiency measurement above.
The $\alpha_N$ parameter was fitted using a purely linear fit to the first four points with the TG filter detuned, see dashed line in Fig.~\ref{fig:result_power}b, yielding $\alpha_N=129(3)\ \text{kHz}/\text{W}/\text{cm}$.
Note that this noise parameter is measured for the TG filter bandwidth of \SI{200}{\pico\meter} (\SI{25}{\giga\hertz}).
Having determined all relevant parameters entering Eq.~\ref{eq:Maring} we can simply compute the theoretical noise curve and compare to the measured data, shown as the solid line in Fig.~\ref{fig:result_power}b.
The agreement is very satisfactory, particularly given that no additional tuning of parameters has been done.

The noise rate at \SI{580}{\nano\meter} was measured with the light coupled into the SMF fiber and using the BP filter and the silicon SPAD.
The BP filter transmitted the noise peak at \SI{580}{\nano\meter}, while suppressing the other noise peaks in the spectrum.
Using the known filter function and the recorded SMF spectrum shown in Fig.~\ref{fig:result_spectrum}b, we estimate that \SI{77}{\percent} of the recorded counts stems from the 580 noise peak.
The resulting noise rate as a function of pump power, shown in Fig.~\ref{fig:result_power}c, was therefore normalized with respect to this fraction, and the transmission coefficient from the output of the waveguide to the detector (including the detector efficiency).

The \SI{580}{\nano\meter} noise rate increases quadratically for low pump powers, as expected for a cascaded SPDC/SFG process.
At higher pump powers, however, the scaling is close to linear, which is due to the saturation of the SFG efficiency for high pump powers.
From Eq.~\ref{eq:Maring} we expect a visible noise rate given by
\begin{align}
R_\text{vis}(P_p)
&=\alpha_N P_p L\, \frac{\eta_\text{max}}{2}\left(1-\frac{\sin\left(2L\sqrt{\eta_\text{n} P_p}\right)}{2L\sqrt{\eta_\text{n} P_p}}\right).
\label{eq:SFGnoise}
\end{align}
We fit the measured noise rate to Eq.~\ref{eq:SFGnoise} with $\alpha_N$ being the only free parameter ($\eta_\text{n}$ and $\eta_\text{max}$ are fixed to the previously obtained values).
The fit is excellent and yields $\alpha_N=\SI{391(16)}{\kilo\hertz\per\watt\per\centi\meter}$, see the solid line in Fig.~\ref{fig:result_power}c.
The $\alpha_N$ parameter is 3 times higher than the value obtained from the telecom noise data, which can be understood by considering the different measurement bandwidths.
Indeed, the telecom noise was measured using the \SI{200}{\pico\meter} TG filter, which is $\num{2.5(2)}$ times narrower than the \SI{500}{\pico\meter} wide SFG dip in the telecom noise (cf.~section \ref{sec:results:spec_analysis}), while the \SI{580}{\nano\meter} noise peak was measured over its full bandwidth.
Without the telecom TG filter we would have expected a telecom noise rate of about $129 \times 2.5 = \SI{323(27)}{\kilo\hertz\per\watt\per\centi\meter}$, which is in reasonable agreement with the observed rate coefficient for the visible noise.
We finally note that in the low-power regime the unnormalized sinc-function in Eq.~\ref{eq:SFGnoise} can be approximated as $\sin(x)/x \approx 1-x^2/6$, resulting in a quadratic power dependence $R_\text{vis}(P_p) = \frac{1}{3} \alpha_N \eta_\text{n} \eta_\text{max}  L^3 P_p^2$.
The dashed line in Fig.~\ref{fig:result_power}c was calculated using this formula with the $\alpha_N$ given for the visible noise rate.

\section{Conclusion}
\label{sec:results:conclusion}

We have presented a device for frequency conversion of photons from \SI{580}{\nano\meter} to the telecom C-band based on the DFG process.
The maximum external conversion efficiency was \SI{46}{\percent}, corresponding to an internal efficiency of \SI{67}{\percent} inside the waveguide.
The noise properties of the device were investigated in detail spectrally as a function of pump power, both in the visible range (\SI{580}{\nano\meter}) and in the telecom range (\SI{1541}{\nano\meter}).
As expected, we find that the main noise mechanism in the telecom range is broadband non-phase-matched SPDC caused by the strong pump laser.
In addition, it is also shown that subsequent SFG conversion of these telecom noise photons has a significant impact on the noise spectrum and the noise rate power dependence.
We clearly identify spectral dips in the telecom noise due to SFG in different spatial modes of the waveguide, and the corresponding noise peaks in the visible spectrum around \SI{580}{\nano\meter}.
The telecom noise rate was $\alpha_N=\SI{129}{\kilo\hertz\per\watt\per\centi\meter}$, measured in a \SI{25}{\giga\hertz} bandwidth.
It is more interesting, however, to express the noise rate in terms of photons per spectro-temporal mode, by dividing $\alpha_N$ by the measurement bandwidth.
We then obtain about $\SI{5e-6}{\per\watt\per\centi\meter}$, which is the probability to generate a noise photon within a spectro-temporal mode given by the signal photon bandwidth.
If the filter bandwidth is larger than the bandwidth of the coverted photons, then the noise probability must be multiplied by the ratio of the filter bandwidth to the photon bandwidth.
For the quantum systems under consideration here (NV$^-$ centers, Eu$^{3+}$ and Pr$^{3+}$ ions), where bandwidths are in the 1 to 10s of MHz range, strong filtering would be required to reach this noise level.
However, given the very low noise per spectro-temporal mode, it is likely that a specific application can be achieved with weaker filtering.

\section*{Funding Information}

This work was financially supported by the European Research Council under AdG project MEC (GA 339198) and the National Swiss Science Foundation (SNSF) under research project no.~172590.

\section*{Acknowledgements}

We thank Nuala Timoney for initial experimental work, Jean Etesse, Alexey Tiranov and Rob Thew for useful discussions, and Claudio Barreiro for technical assistance.

\bibliography{qmcommon}

\end{document}